\newcommand{\un}{~\mathrm}

\documentclass{elsart}

\usepackage[T1]{fontenc}% pour utiliser les caractères français
\usepackage{graphicx}% Include figure files
\usepackage{dcolumn}% Align table columns on decimal point
\usepackage{bm}% bold math
\usepackage{amssymb}

\begin{document}

\begin{frontmatter}

% Title, authors and addresses

% use the thanksref command within \title, \author or \address for footnotes;
% use the corauthref command within \author for corresponding author footnotes;
% use the ead command for the email address,
% and the form \ead[url] for the home page:
% \title{Title\thanksref{label1}}
% \thanks[label1]{}
% \author{Name\corauthref{cor1}\thanksref{label2}}
% \ead{email address}
% \ead[url]{home page}
% \thanks[label2]{}
% \corauth[cor1]{}
% \address{Address\thanksref{label3}}
% \thanks[label3]{}

\title{Fracture of glassy materials as detected by {\em real-time} Atomic Force Microscopy  (AFM) experiments.}

\author[ldv]{F. Célarié}
\author[cea]{S. Prades}
\author[ldv,cea]{D. Bonamy}
\author[ldv]{A. Dickelé}
\author[ldv]{L. Ferrero}
\author[cea]{E. Bouchaud}
\author[cea]{C. Guillot}
\author[ldv]{C. Marlière\corauthref{cor1}}

\address[ldv]{Laboratoire des Verres - UMR CNRS-UM2 5587, Université Montpellier 2, C.C. 69 - Place Bataillon, F-34095 Montpellier Cedex 5 - France}
\address[cea]{Service de Physique et Chimie des Surfaces et Interfaces, DSM/DRECAM/SPCSI, CEA Saclay, F-91191 Gif sur Yvette - France}

\corauth[cor1]{Phone: +33 4 67 14 37 06, Fax: +33 4 67 14 33 98}
\ead{marliere@ldv.univ-montp2.fr}

\begin{abstract}

We have studied the low speed fracture regime for different glassy materials with variable but controlled length scales of heterogeneity in a carefully mastered surrounding atmosphere. By using optical and atomic force (AFM) microscopy techniques we tracked in real-time the crack tip propagation at the nanometer scale on a wide velocity range ($10^{-3} - 10^{-10}\un{m/s}$ and below). The influence of the heterogeneities on this velocity is presented and discussed. Our experiments revealed also -~for the first time~- that the crack advance proceeds through nucleation, growth and coalescence of nanometric damage cavities {\em inside the amorphous phase}, which generate large velocity fluctuations. The implications of the existence of such a nano-ductile fracture mode in glass are discussed.

\end{abstract}

\end{frontmatter}

% main text

\section{Introduction}

Fracture of brittle materials remains far from being understood. One of the main problems comes from the fact that cracks imply several length scales \cite{Mcclintock66} from the atomic to the macroscopic scale, as revealed for instance by the fracture surfaces morphology which presents two regimes, a low length scales regime with a roughness exponent $\zeta=0.5$ and a large length scales regime with  a roughness exponent $\zeta\simeq 0.8$ \cite{Daguier95,Bouchaud97} separated by an intermediate scale crossover length $\xi_c$ ranging from $10$ to $30\un{nm}$. We propose here to investigate the quasi-static fracture regime, observed in stress corrosion, at these scales. In this context, we have designed an experimental setup, based on an Atomic Force Microscope (AFM), which allows us to track in real time the crack tip propagation in more or less devitrified vitroceramics for velocities ranging from $10^{-3}\un{m/s}$ to $10^{-12}\un{m/s}$ at nanometric length scales. Variations of the crack tip velocity with the stress intensity factor have been measured for amorphous and partially devitrified materials. The influence of the heterogeneities is enlightened and explained. Moreover, both samples present important velocity fluctuations with standard deviation of the same order as the mean lowest velocities. These fluctuations are shown to be related to the presence of nanoscale damage cavities ahead of the crack tip {\em within the amorphous phase}. These look somewhat similar to the cavities observed ahead of the crack tip during the ductile fracture of a metallic alloy, which may explain the similarity of fracture morphologies observed in glass  and metallic alloys \cite{Daguier95}.

\section{Experimental setup}

The experimental setup is illustrated in Fig.~\ref{fig1}. All the experiments are performed at a constant temperature of $22.0\pm 0.5~^\circ\mathrm{C}$ in a leak-proof chamber under an atmosphere composed of pure nitrogen and water vapour at a relative humidity level of $42\pm 1\%$ after preliminary out-gassing. We have chosen to investigate fractures in aluminosilicate vitroceramics that can be more and less devitrified by controlling the thermal processing \cite{Marliere02}. Both the size and the volume fraction of crystalline grains are evaluated by imaging the samples surface after a stay in fluorhydric acid (concentration ranging from $0.4\%$ to $2\%$) for $30\un{s}$, which dissolves faster the amorphous phase than the crystalline one~\cite{Dickele02}.

Fracture is performed on DCDC~\cite{He95} (Double Cleavage Drilled Compression) specimens: Parallelepipedic ($4\times4\times40\un{mm}^3$) samples are designed with a cylindrical hole (radius $a=0.5\un{mm}$) drilled in the center and perpendicularly to the $4\times40\un{mm}^2$ surface (see Fig.~\ref{fig1}a). The hole axis defines the $z$-direction. The $x$-axis (resp. $y$-axis) is parallel to the $40\un{mm}$ (resp. $4\un{mm}$) side of the $4\times40\un{mm}^2$ surfaces. A thermal treatment ($660^\circ\mathrm{C}$) is applied to obtain pure amorphous material free of residual stresses. Higher temperature (two stages of $10\un{mn}$ respectively at $750~^\circ\mathrm{C}$ and $900~^\circ\mathrm{C}$) are applied to obtain the partially devitrified samples \cite{Marliere02}. In both cases, the $4\times40\un{mm}^2$ surfaces are then optically polished (the RMS roughness is around $0.25\un{nm}$ for a $10\times10~\mu\mathrm{m}^2$ scan size). A compressive load is applied perpendicularly to the $4\times4\un{mm}^2$ surfaces. The external stress $\sigma$ is gradually increased by the slow constant displacement ($0.02\un{mm/min}$) of the jaws of the compressive machine (Fig.~\ref{fig1}b). Once the two cracks (symmetrically to the hole axis) are initiated, the displacement of the jaws is stopped. The crack then propagates along the $x$-axis in the symmetry plane of the sample parallel to the ($x$,$z$) plane. In this geometry, the stress intensity factor $K_I$ is given by \cite{He95}: $K_I=\sigma\sqrt{a}/(0.375c/a+2)$, where $c$ is the crack length (Fig.~\ref{fig1}a).

The crack motion within the ($x$,$y$) surface is monitored by our experimental system combining optical microscopy and AFM. Optical image processing gives the position of the crack tip and consequently the ``instantaneous" velocity for $v$ ranging from $10^{-6}$ to $10^{-9}\un{m.s}^{-1}$. By AFM measurements -~performed in a high amplitude resonant mode (``tapping" mode)~-, one probes the crack tip neighbourhood at magnifications ranging from $75\times75\un{nm}^2$ to $5\times5~\mu\mathrm{m}^2$ and the crack tip motion at velocities ranging from $10^{-9}$ to $10^{-12}\un{m.s}^{-1}$.

\section{Results}

At the very first moments, the crack propagates very quickly. But as the crack length $c$ increases, $K_I$ decreases. When $K_I$ becomes smaller than the fracture toughness $K_{Ic}$ (i.e. in the stress corrosion regime), the crack motion is slow enough to be monitored by our experimental system. Figure~\ref{fig2} shows the variation of the velocity $v$ as a function of the stress intensity factor $K_I$ for both the amorphous and the partially devitrified specimens. The exponential behaviour is compatible with stress enhanced activated process models \cite{Wiederhorn67,Wiederhorn70}:

\begin{equation}
\begin{array}{ll}
& v(K_I)=v_0\exp(-\alpha (K_I-K^*_{I}))/kT)\\
& v_0=\nu_0 a_0\exp(-\Delta {F^*}/kT)
\end{array}
\label{equ1}
\end{equation}

\noindent where $\nu_0$ is a characteristic lattice frequency, $a_0$ a characteristic atom spacing, $\Delta F^*$ a quiescent adsorption/desorption activation energy, $\alpha$ an activation area and $K^*_{I}$ a characteristic value of the stress intensity factor~\cite{Lawn93}. All these parameters depend {\em a priori} on the chemical and mechanical properties of the material and are expected to be different in the two samples.

A constant positive shift in the $v(K_I)$ curves in Fig.~\ref{fig2} can be observed for the devitrified specimen.
The velocities measured for the partially devitrified sample are shifted toward positive $K_I$ when compared to the data relative to the amorphous sample. To understand this shift, we probe the surface crack path at sub-micrometric scales (Fig.~\ref{fig3}): In the partially devitrified specimen, the crystalline germs deflect the crack (Fig.~\ref{fig3}b). As the crack tip keeps going through the amorphous phase, the chemical constants $\nu_0$, $a_0$, $\Delta F^*$ and $\alpha$ in Eq.~\ref{equ1} -~and consequently the slope of the semilogarithmic curves $v(K_I)$~- are the same in both the amorphous and the partially devitrified specimens. However, the deflections of the crack by the crystals induce mode II and mode III components in the local stress intensity factor, which toughens the material~\cite{Faber83}. Consequently, $K_{Ic}$ and then  $K^*_{I}$ is larger in the devitrified sample, which shifts the $v(K_I)$ curve. A similar effect of the size of heterogeneities has been observed for the fatigue of metallic alloys \cite{Ducourthial01}.

The lowest velocities shown in Fig.~\ref{fig2} exhibit important fluctuations -~of the order of the average velocity~- for {\em both} specimens. Consequently,  they cannot be related to the crystalline heterogeneities but are more likely inherent to the amorphous phase.
To understand their origin, we probed the neighbourhood of the crack tip at the nanometer scale in the amorphous specimen (Fig.~\ref{fig4}). This clearly reveals cavities of typically $20\un{nm}$ in length and $5\un{nm}$ in width ahead of the crack tip (Fig.~\ref{fig4}a). These cavities grow with time (Fig.~\ref{fig4}b) until they coalesce (Fig.~\ref{fig4}c). At these nanometric scales, the crack front does not propagate regularly, but intermittently through the merging of the nano-scale cavities, which explains the large fluctuations observed for the lowest velocitites (Fig.~\ref{fig2}). These spots were shown to be nanoscale damage cavities~\cite{Celarie02} similar to the ones commonly observed at the micrometric scale in metallic alloys~\cite{Pineau95,Bouchaud99}, which may explain the departure from linear elasticity observed in the vicinity of a crack tip~\cite{Guilloteau96,Henaux00} in glass as well as the striking similarity of the morphologies of fracture surfaces of glass and metallic alloys at different length scales~\cite{Daguier95,Bouchaud97}. Such a nanoscale ductile fracture mode -~ first observed in Molecular Dynamics simulations~\cite{Nakano99,Vashishta99,Campbell99,VanBrutzel02} has been also evidenced in Silica ~\cite{Celarie02}, and seems to be independent of the precise chemical composition of the  glass. The following scenario is then proposed: The lowest density zones behave as stress concentrators and grow under the stress imposed by the presence of the main crack to give birth to the cavities which are actually observed.

\section{Conclusion}

Quasi-static fracture in glassy materials has been studied {\em in real time} at sub-micrometric length scales in a wide velocity range. The crack velocities are measured as a function of the stress intensity factor in both an amorphous material and a partially devitrified one. The presence of crystalline hetogeneities is shown to toughen the glass. Moreover, important fluctuations are reported for the lowest velocities. They are conjectured to be related to the presence of nano-scale damage cavities observed in real time ahead of the crack tip. The implications of such a ductile fracture mode on the morphology of the fracture surfaces have been discussed. However it should be noted that our investigations are performed on the sample surface while the fracture surface morphology is related to the fracture of bulk. Consequently, it would be interesting to investigate the 3D distribution of damage cavities. Work in this direction is currently in progress.

\newpage

\begin{figure}
\centering
\caption{ experimental setup. (a): Sketch of the DCDC geometry (b): Picture of the experiment.} 
\label{fig1}
\end{figure}

\newpage

\begin{figure}
\centering
\caption{Variation of the crack tip velocity $v$ versus the stress intensity factor $K_I$. Open circles (resp.  Black triangles) correspond to optical measurements (resp. AFM measurements). The error bars correspond to the standard deviation on the velocity for a fixed value of the stress intensity factor $K_I$. For the lowest velocities $v$, the fluctuations are of the order of the average velocity.} 
\label{fig2}
\end{figure}

\newpage

\begin{figure}
\centering
\caption{Crack path in (a) the amorphous specimen and (b) the partially devitrified one. Frames are phase images sensitive to local mechanical properties, which allows to distinguish the crystalline heterogeneities. In the partially devitrified sample, the crystals deflect the crack which propagates within the amorphous phase.} 
\label{fig3}
\end{figure}

\newpage

\begin{figure}
\centering
\caption{Sequence of successive topographic AFM frames showing the crack propagation at the surface of the amorphous specimen. The crack front propagates from the left to the right ($x$-direction) with an average velocity $v$ close to $10^{-11}\un{m/s}$. (a): evidence of nanometric damage cavities ahead of the crack tip. (b): growth of the cavities. (c): the crack is propagating via the coalescence of all the cavities. The data presented here are obtained for  $K_I=0.43\mathrm{MPa.m}^{1/2}$ and $v=3.10^{-11}\un{m/s}$}
\label{fig4}
\end{figure}

\end{document}